\documentclass[11pt,3p,review,authoryear]{elsarticle}

\usepackage{amssymb}
\usepackage{amsmath}
\usepackage{array}
\usepackage[table]{xcolor}
\usepackage{colortbl}
\usepackage{graphicx} 
\usepackage{subcaption} 

\definecolor{lightgreen}{RGB}{192, 255, 192}
\definecolor{mediumgreen}{RGB}{110, 228, 110}
\definecolor{mediumgreen2}{RGB}{70, 198, 70}
\definecolor{darkgreen}{RGB}{30, 150, 30}

\journal{International Journal of Forecasting}
\biboptions{longnamesfirst}

\begin{document}

\begin{frontmatter}

\title{SolNet: Open-source deep learning models for photovoltaic power forecasting across the globe}

\author[1]{Joris Depoortere}
\cortext[cor]{Corresponding author}
\ead{joris.depoortere@kuleuven.be}
\author[1]{Johan Driesen}
\author[1]{Johan Suykens}
\author[1]{Hussain Syed Kazmi}
\affiliation[1]{organization={ESAT, KU Leuven},
            addressline={Kasteelpark Arenberg 10}, 
            city={Leuven},
            postcode={3001}, 
            country={Belgium}}

\begin{abstract}
Deep learning models have gained increasing prominence in recent years in solar photovoltaic (PV) forecasting. One drawback of these models is that they require a lot of high-quality data to perform well. This is often infeasible in practice, due to poor measurement infrastructure in legacy systems and the rapid build-up of new solar systems across the world. This paper proposes SolNet: a novel, general-purpose, multivariate solar power forecaster, which addresses these challenges by using a two-step forecasting pipeline which incorporates transfer learning from abundant synthetic data generated from PVGIS, before fine-tuning on observational data. 

Using actual production data from hundreds of sites in the Netherlands, Australia and Belgium, we show that SolNet improves forecasting performance over data-scarce settings as well as baseline models. We find transfer learning benefits to be the strongest when only limited observational data is available. At the same time we provide several guidelines and considerations for transfer learning practitioners, as our results show that weather data, seasonal patterns and possible misspecification in source location can have a major impact on the results. The SolNet models created in this way are applicable for any land-based solar photovoltaic system across the planet to obtain improved forecasting capabilities.
\end{abstract}


\end{frontmatter}

\section{Introduction}
Renewable energy sources (RES) are seen as a critical component of the energy transition necessary for climate change mitigation \citep{panwar2011role}. However, the actual generation from these RES depends on local weather conditions and is known for its variability and intermittency. As such, expanding generation through RES introduces additional uncertainty in the power system operation, necessitating sophisticated generation forecasts on hour-ahead to day-ahead levels \citep{notton2018intermittent}. 

However, with rapidly growing solar and wind installed capacity, both the average and worst case forecast errors have also rapidly grown. Solar power forecasting is especially problematic, with the worst-case day-ahead generation forecast error (mean absolute error calculated on hourly forecasts) having grown from around 3 GW to over 13 GW in 16 European countries in just the past few years \citep{kazmi2022good}. Moreover, the proliferation of rooftop systems in the distribution grid further complicates grid operation by introducing over-voltage and reverse power flow concerns \citep{steffel2012integrating}. Additionally, unlike for large solar plants where information is systematically logged, forecasting power generation from these small scale distributed solar systems is further complicated by the lack of relevant meta-data (e.g. system size, orientation etc.) and supporting forecasts for related exogenous factors (e.g. ambient conditions, cloud cover, etc.).

Motivated by these concerns, solar power forecasting has seen significant progress as a field in recent years \citep{hong2020energy}. Inspired by the state-of-the-art performance of deep learning in other domains such as computer vision and natural language processing, several researchers have also utilized these methods to build solar forecasters \citep{kumari2021deep}. The results of naively applying deep learning in forecasting, however, are mixed. Even though no solar-specific benchmarks are available, Makridakis et al. discuss the apparent under-performance of pure deep learning models, compared to hybrid or pure statistical models after analysing the results of the M4 forecasting competition \citep{makridakis2018m4}, a key competition in the field of time series forecasting. 
The Great Energy Predictor (GEP-III) competition, focused on forecasting building energy demand, has likewise resulted in similar conclusions about simpler methods such as gradient boosting often outperforming deep learning methods \citep{miller2020ashrae}. 

A major reason for deep learning underperforming simpler models in many forecasting tasks is its requirement for enormous amounts of data to generalize well. On the contrary, conventional time-series models (e.g. those from the SARIMAX family), and simpler machine learning based models (e.g. those built on tree-based methods) can generalize relatively well from limited amounts of data. Consequently, these models do not suffer from the fact that sufficient historical data on solar operation or exogenous factors is often not readily available in practice. 

\subsection{State-of-the-art}

In recent years, transfer learning has emerged as a key component in workflows aimed at improving the sample complexity of deep learning methods (i.e. the amount of data required to generalise well) \citep{weiss2016survey}. This has been well-documented in several domains including computer vision (CV) \citep{gupta2022deep}, natural language processing  (NLP) \citep{ruder-etal-2019-transfer}, and reinforcement learning tasks \citep{zhu2023transfer}. In this workflow, rather than training a model directly on (sparse) observational data, a source model is first trained on a large corpus of data (derived from a known source domain). This source model is then fine-tuned using sparse or limited observational data (derived from the target domain of interest). Foregoing the initial pre-training phase (i.e. utilizing data from the source domain) leads to poor generalisation due to data sparsity in practice. The closer the source domain is to the target domain, the better the results of transfer learning.

Despite the pressing needs of the sector and the potential of transfer learning, practical examples of applying transfer learning to solar forecasting are still few and far between \citep{peirelinck2022transfer}.
For instance, Zhou et al. have recently applied transfer learning to PV power forecasting used a 3-layer LSTM architecture to predict PV power output, where the entire network was first trained on the source domain (solar irradiance), with the last layer fine-tuned on the target data (PV output) \citep{zhou2020transfer}. Interestingly, even though the source and target domain are somewhat different, the authors report improved predictive accuracy. Even more recently, \cite{abubakr2022unleashing} have used transfer learning to predict solar irradiance at solar farms with limited data by using the abundant data available at one site. The model architecture was a Recurrent Neural Network (RNN), using historical solar irradiance data with up to four exogenous weather variables. This results in (slightly) better performance when using transfer learning compared to training the model directly on the limited data. The relatively small improvement in performance could well be due to the fact that the prediction time horizon is only one hour in the future.

These studies are emblematic of the issues hampering adoption of transfer learning in the energy field. Transferring from data observed at other, similar sites is not fully generalisable, i.e. it introduces a new requirement during the model's training process. One way to avoid this issue is by using synthetic data to perform the initial source model creation. However, despite early work by Zhang et al. using a generative adversarial network (GAN) to create synthetic time series of solar generation 
\cite{zhang2018generative}, this has not been pursued further. Using data-driven generative models also suffer from some of the same issues: the GAN (or other trained generative model) requires previously collected datasets to generalize well, which is the actual issue to begin with.

The studies published so far also suffer from several limitations in terms of their evaluation robustness and transfer applicability. They typically utilise very limited testing or future look-ahead periods, raising questions about the statistical significance of the reported results. The seasonal variations of solar production over the year limits the generalisability of these results even more. Furthermore, these studies typically report improvements only in asymptotic performance, i.e. do not consider the influence of increasing data on model accuracy, a critical concern in transfer learning studies. A more complete evaluation of transfer learning would look at differing amounts of training data, up until the initial performance, with no target data to fine-tune on. We elaborate on this further in the methodology section.

Finally, arguably the most important reason behind the wide-spread adoption of source models in computer vision and natural language processing has been their wide-spread availability to researchers and practitioners alike. The few source models that have been trained by researchers for energy or solar applications are all invariably closed-source and inaccessible, having been trained from scratch themselves \citep{HU201683, QURESHI2017742, zhou2020transfer, abubakr2022unleashing}. This largely defeats the purpose of transfer learning.

\subsection{Contributions}

Consequently, in this paper, we make two major contributions to the existing literature. First, we propose a novel way to forecast residential PV using transfer learning. This results in SolNet, a unique open-source deep learning model generator for solar power forecasting for any land-based site on the planet. The proposed models, made available for use through a Python package (GitHub link in section \ref{conclusion}), rely only on API access to PVGIS \citep{HULD20121803} and Open-Meteo \citep{Zippenfenig_Open-Meteo}. PVGIS is used to generate PV production data for several variations of weather conditions and panel specifications (e.g. tilt, azimuth, etc.). Open-Meteo provides historical weather data, dating back for over 80 years. Using this large corpus of data for learning removes the need for data collection for end-users, makes the transfer approach more realistic, and applicable to largely any land-based location on the planet. Once the source model has been trained, it can be fine-tuned with observational data where it exists. 

\begin{figure}[!t]
\centering
\includegraphics[width=4in]{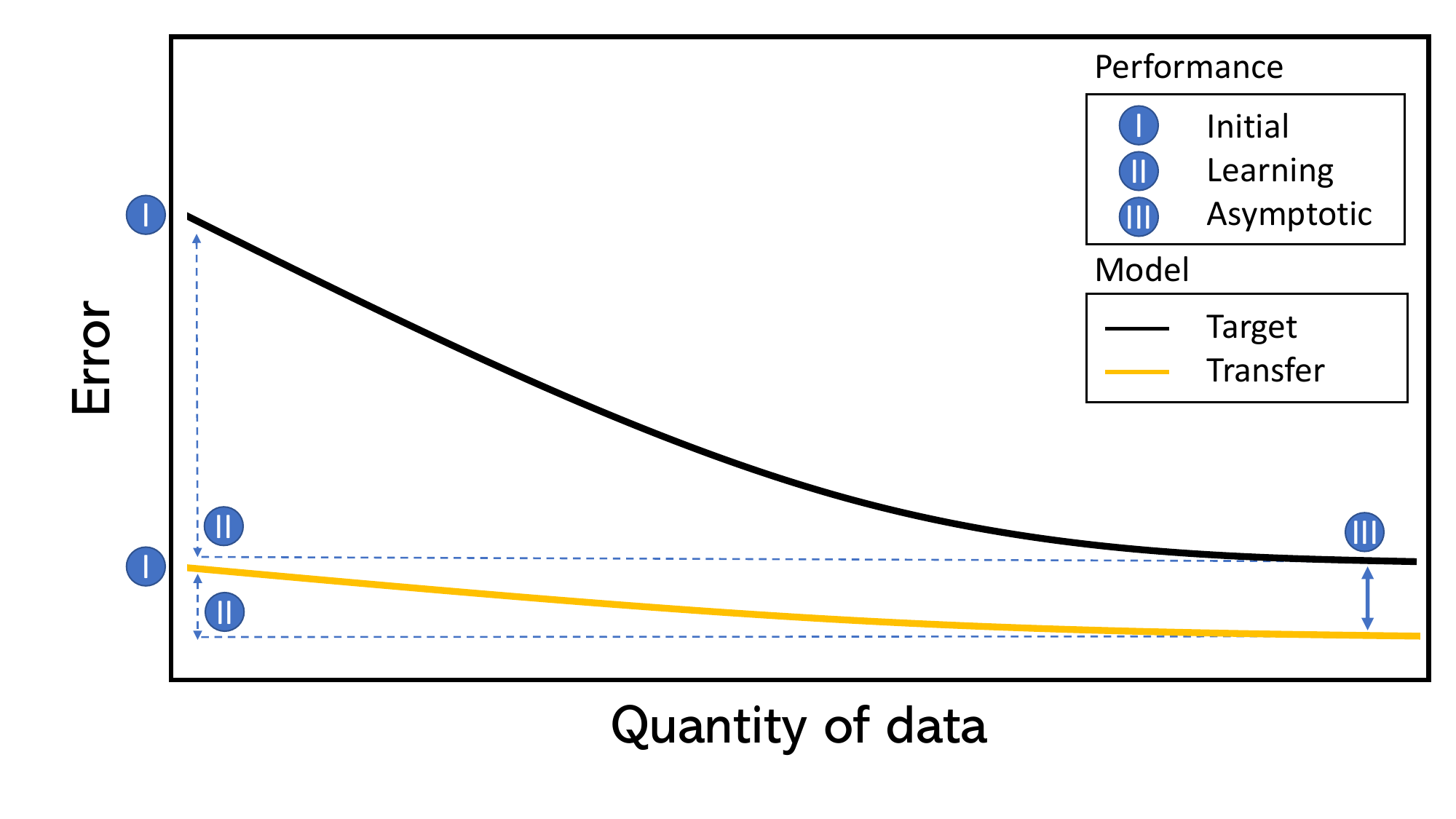}
\caption{Complete assessment of transfer performance. The initial performance measures performance of a model without any training in the domain of interest (the target domain). The learning performance indicates the performance gain we can expect over time as more data to train on becomes available. Finally, the asymptotic performance is the 'permanent' performance differential between the models when sufficient data is available. The shape of the figure is indicative of our hypothesised findings.}
\label{fig_performance}
\end{figure}

Second, we do a thorough deep-dive of various transfer learning scenarios, providing guidance on when and how to use it successfully in solar power generation scenario's. We look at three performance metrics for better understanding of the value of transfer: initial performance, learning improvement, and asymptotic performance. Figure \ref{fig_performance} visualises our hypothesis of the expected performance of the transfer learning. Additionally, we look at (1) the importance of weather data, (2) the impact of the yearly seasonality on transfer learning out-performance, and (3) how misspecification of the source domain can impact the performance of transfer learning. 

\subsection{Organization}

The remainder of this paper is structured as follows: Section 2 discusses the forecasting exercise, the models utilised for forecasting, and the transfer learning methodology. Section 3 elaborates on the data used in this transfer learning process, including the source (PVGIS and Open-Meteo) and target (the actual site locations and DWD ICON EU) domains, and how they are prepared for the analysis. Section 4 provides detailed results from applying the proposed methodology, while Sections 5 and 6 conclude the paper, going over the most important findings as well as important future research directions.

\section{Methodology}

We first discuss the design of the forecasting exercise, as well as the benchmark used for comparison purposes of our transfer approach. Next, we discuss the architecture of our chosen neural network, LSTM (long short term memory), a type of recurrent neural network which can learn temporal associations. Then we elaborate on the transfer learning methodology applied to the LSTM models. Finally, we talk about the metrics used for the evaluation.

\subsection{The forecasting exercise and baseline approach}

Our forecasting exercise presents two cases, a univariate and multivariate case. In the univariate case we only use 24 lagged values of the target (PV power), and cyclically encoded values of the hour of the day. While we are aware that these are additional covariates, we still consider this case as univariate, as no external information is required for the cyclical encoding, we only rely on the hour of the day for these covariates. In the multivariate case we include several weather parameters as future covariates, as well as the 24 lagged values of PV power. The reader is referred to section \ref{weather_data_section} for more information on the weather data.

For this paper we focused on the case of forecasting the next 24 hours, from 00:00 to 23:00 with no gap between inference time and the forecast period. This is a simple reference case to showcase the performance of SolNet. The main reason for choosing this option is that, when using weather forecast data (which we get from the Deutscher Wetterdienst's Icosahedral Nonhydrostatic EU model (DWD ICON EU) via Open Climate Fix \citep{open_climate_fix_2023}), we have to decide on a reference timing. Historical weather forecasts, when and if they are available, typically offer the option to choose from four different reference time slots: 00:00, 06:00, 12:00 and 18:00. We limited ourselves to the 00:00 data, as otherwise the dimensionality of the data would have become too large to work with\footnote{Note that DWD forecasts are publicly available several hours after the reference time of 00:00, meaning that a user should either use 2 day-ahead forecasts or a different source for real-life application. We stick to DWD because it is one of the few public sources of historcal weather forecasts.}.

To compare the performance of our deep learning model, we use a Naive seasonal model (K=24). This model takes the corresponding hour of the previous day as the prediction for the next day. So, if the actual power generated at 14:00 on day $d$ was 0.45 kW, the forecast value for 14:00 on day $d+1$ will also be 0.45 kW. This method provides us with a realistic baseline, given the diurnal pattern of solar power. We will refer to this method as the 'naive' or 'naive seasonal' method for the remainder of this paper. 

\subsection{LSTM as a forecast model}

The deep learning architecture used in our experiments, both with and without transfer learning, is the Long Short-term Memory (LSTM) model \citep{hochreiter1997long}. An LSTM model consists of several identical units. These units have several gates that remember (or forget) information over time, allowing the model to retain a form of memory. This enables LSTMs to learn long term patterns in time series data. All the gates in an LSTM are made from the fundamental building blocks of neural networks: weights, biases and non-linearities in the form of activation functions (e.g. sigmoid or tanh functions). These weights and biases are the parameters that are learned during the training process. The below formulas provide an overview of the internals of the LSTM unit. 

\begin{flalign}
& \text{Input Gate:} \quad && i_t = \sigma(W_{xi}x_t + W_{hi}h_{t-1} + b_{i1}) \times \tanh(W_{xi}x_t + W_{hi}h_{t-1} + b_{i2}) & \\
& \text{Forget Gate:} \quad && f_t = \sigma(W_{xf}x_t + W_{hf}h_{t-1} + b_f) & \\
& \text{Output Gate:} \quad && o_t = \sigma(W_{xo}x_t + W_{ho}h_{t-1} + b_o) & \\
& \text{Cell State:} \quad && c_t = f_t c_{t-1} + i_t & \\
& \text{Hidden State:} \quad && h_t = o_t \tanh(c_t) &
\end{flalign}
\begin{align*}
\text{where:} & \\
W_{xn} & \text{: are the weights for the input} \\    
W_{hn} & \text{: are the weights for the hidden state} \\
b_n & \text{: are the biases} \\
h_{t-1} & \text{: is the hidden state at time $t - 1$} \\
c_{t-1} & \text{: is the cell state at time $t - 1$} \\
\end{align*}

Key to the LSTM are the cell state (the long-term memory) and the hidden state (the short-term memory). These two states get updated based on the states at the previous time step and the input associated with the current time step, after they pass through several gates. These gates manage what gets passed on to the next states (and what gets output).

The relative simplicity of the LSTM model, existing out of \textit{N} layers with identical building blocks, taking (scaled) inputs and running them through several activation functions after applying weights and biases to them, makes it a good candidate for transfer learning. The fine-tuning options are (relatively) limited compared to some modern architectures, and the only prerequisite of using the model in both the target domain and the source domain, is that the inputs to the models are of equal shapes (i.e. without introducing additional complexity). In the univariate case, this prerequisite is easily fulfilled as the model only requires the user to have \textit{K} hours of historical PV output data available, with \textit{K} being the number of historical lags used to make the forecast. For the purpose of this research we limit this to $K = 24$. From trial and error we noted that adding more lags did not add significantly to training performance, whilst increasing the training time by a significant margin. Due to this research using over a hundred test sites with many configurations, this additional training time was not practical to account for. 

The final architecture of the LSTM used in our experiments was found after doing a grid search for the best hyper-parameters (layers, neurons per layer, dropout percentage), and includes 3 layers, each consisting of 400 LSTM units, with one more layer consisting of 24 linear units. These 24 units form the outputs, one for each hour of the day. Other hyper-parameters include the optimizer used during the back-propagation process (ADAM), the learning rate ($10^{-4}$ in the base case and $10^{-6}$ in the fine-tuning case, see later in this section), a dropout rate to avoid overfitting (50\% in our case), and batch size (32). These hyperparameters (optimizer, learning rate, dropout rate and batch size) were chosen based on trial and error and best practices from published literature. 

\subsection{Transfer learning}

With the deep learning architecture in place, it is possible to explore the effect of transfer learning on solar forecasting in a principled manner. More concretely, the LSTM can be pre-trained on source data, after which it can be fine-tuned using target data to achieve potential improvements over the target model.
Transfer learning, formally defined as improving the learning of a target predictive function on the learning task, $\mathcal{F}_{T}(\cdot)$ in the target domain, $\mathcal{D}_{T}$, using knowledge of $\mathcal{D}_{S}$ and $\mathcal{T}_{S}$ (the source domain and the source task respectively) \citep{pan2010survey}. While $\mathcal{D}_{T} \neq \mathcal{D}_{S}$ and $\mathcal{T}_{T} \neq \mathcal{T}_{S}$, we do assume that, the more similar the two are, the easier the transfer learning task will be. In our case, we keep the source domain as closely related to the target domain as possible by mimicking the location and the PV set-up of our target domain data, in our source domain, but we also provide results for what happens when this condition does not hold (see section \ref{chap:quan_and_qual}). This is in addition to the inherent misalignment between simulated PVGIS data and observed data. 

Concretely, in addition to directly using the pre-trained source model for the target task of solar PV prediction, there are a few different algorithmic ways in which transfer learning can be applied to the LSTM networks:

\begin{enumerate}
\item \textbf{Direct application.} The pre-trained source model can directly be applied in the target domain without any adjustments to the model parameters. 
\item \textbf{Finetuning.} The source LSTM can be fine-tuned to the target domain. We keep the architecture intact, and run several additional training iterations using the (limited) target domain data. We can avoid model overfitting on the limited target data by lowering the learning rate, thereby making sure that the model only makes slight adjustments to the previously trained weights. Another popular method to fine-tune deep neural networks is by ``freezing'' the parameters of certain units or entire layers (see e.g. \cite{9070676}). In this case, the optimiser can no longer update the parameters in these layers during the optimisation process, but is free to update the remaining parameters.
\end{enumerate}

For our experiments, we combine both finetuning strategies: we freeze the initial LSTM layer, and lower the learning rate of the subsequent layers by a factor of 100, when retraining on the target domain data. This ensures that the fine-tuning process does not completely over-write the learning from the source domain. We also look at direct application of the source models. This basically represents the case where we only evaluate the initial performance (zero-shot learning). 

\subsection{Model evaluation}

Performance is measured by way of the root mean squared error (RMSE), in accordance with \citep{yang2020verification}, since it penalizes large errors more heavily. This is important to correctly predict the timing and size of the daily peak in output. Another reason to focus on the RMSE is its ubiquity in solar forecasting literature, which makes it easier to compare across studies. In appendix we also include the mean absolute error (MAE) and mean bias error (MBE) for each of the experiments, to provide a more complete picture of model performance. The MAE focuses more on smaller errors compared to the RMSE and the MBE gives insights towards the bias of the models.

\begin{figure}[!t]
\centering
\includegraphics[width=4in]{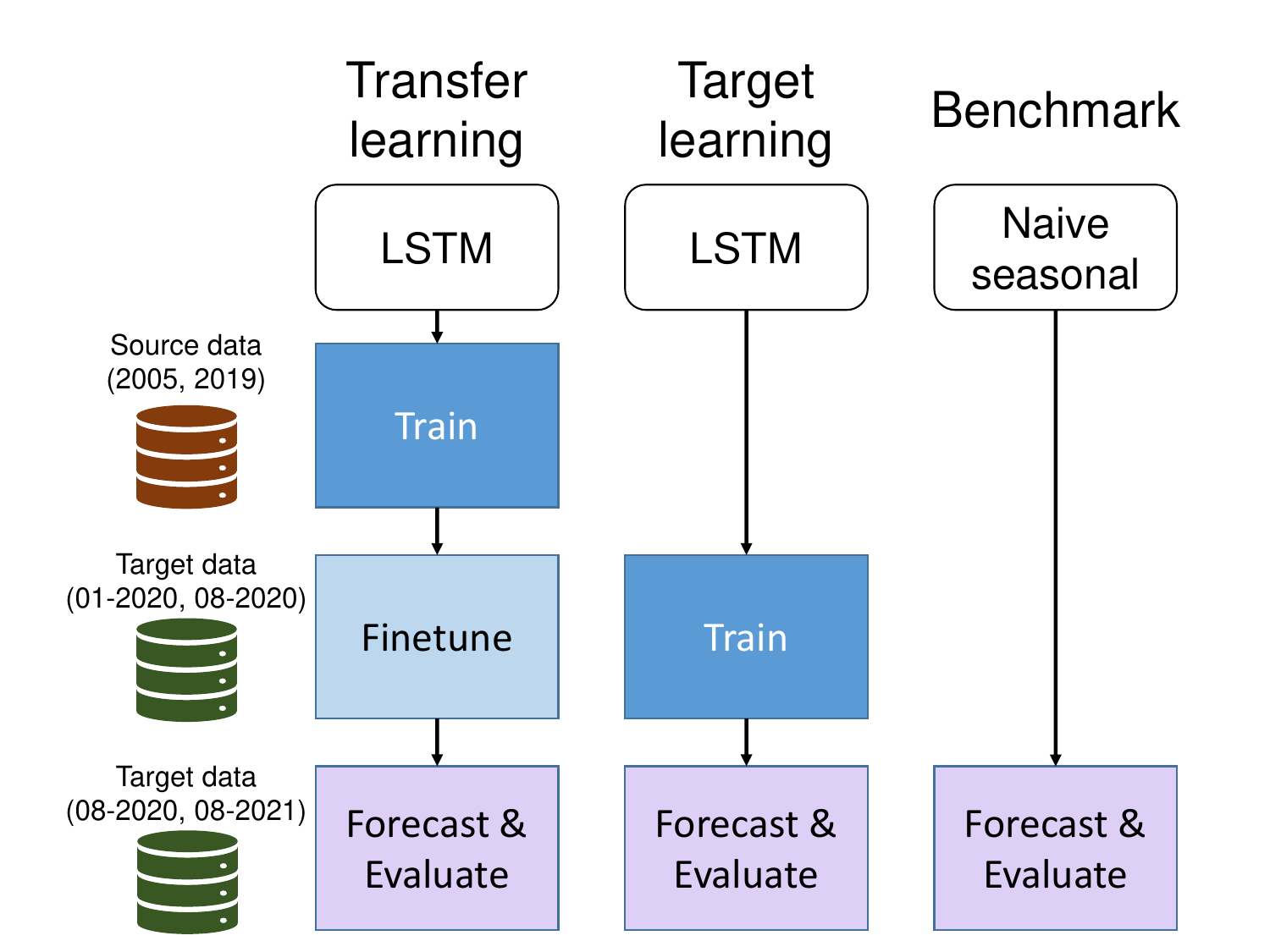}
\caption{The workflow of the experiments. In the transfer learning case we train in the (synthetic) source domain before finetuning in the target domain. In direct (i.e. target) learning, we train directly on the target domain. The benchmark is a persistence model and does not require any training. The periods on the left are specifically those of the NL case, with target data starting in 2020 and ending in August 2021.}
\label{fig_workflow}
\end{figure}

In all cases, a full year of target data is kept aside for final evaluation. 80\% of the remaining data is used for training and 20\% is used to validate the model performance and pick the best performing iteration of the model. The complete workflow of the experiments is presented in figure \ref{fig_workflow}. The first column shows the workflow for a source trained model. The middle column shows the process for a target trained model, skipping the source step and training directly on the target domain instead of finetuning with a lower learning rate and frozen layers. The final column shows the naive seasonal model, which takes the values of the day before as the predictions, requiring no training.

\section{Data}

Two types of data are required for this research: (1) the target domain data which has been observed in physical sites of interest but is often limited in quantity (and perhaps quality as well), and (2) the source domain data, which is available in ample amounts, but has actually been synthetically generated from a (similar) setup. First, we briefly discuss the target domain datasets, i.e. PV installations from the Netherlands, Australia, and Belgium, with only around 1.5 to 2 years of data per site. Second, we discuss the synthetic source domain used for transfer learning. Finally, we discuss the potential to add weather input features, as well as any necessary pre-processing steps.

\subsection{Target domain}

\subsubsection{Residential buildings in the Netherlands}

For the Netherlands we have data from 40 individual household PV systems spread across the country. The data comes at a temporal resolution of 15 minutes (which we resample to an hourly resolution in accordance with the source domain). The Dutch dataset has all the required relevant site-specific meta-data available to further fine-tune our synthetic data, including the tilt, the azimuth, the peak power, and the approximate location of the PV systems (zip code and city). From these location parameters we derive an approximate latitude and longitude. The peak power of the systems varies between 1.2 kWp and 3.1 kWp, with most systems around 2.5 kWp.

For all 40 locations, the full year of 2020 is available as well as 2021 up until August. We also have a large part of 2019 available for most locations, but unfortunately our weather forecast data is limited to 2020 and 2021, so we cannot go back further for target locations if we want include weather variables. As mentioned, a full year of data is kept aside for evaluation of the final models. Which will thus consist of a part of 2021 and a part of 2020. For the univariate analysis, this leaves us with a full year (for most locations) to evaluate performance, but for the weather analysis we only have January 2020 up until August 2020. This leaves little data for training and validation in the target domain, but this fits well with the core premise of the research that there is often only a (very) limited amount of data available in the target domain before forecasts need to be made. 

\subsubsection{Residential buildings in Australia}

We also utilise solar data from 300 Australian homes with rooftop solar systems spanning the period of 1 July 2011 to 30 June 2013. The solar systems have a mean size of 1.68 kWp and are all located in New South Wales (specifically in zip codes between 20 and 23). No meta-data is available for these locations other than the size of the system and the zip code. For this dataset we will only do univariate forecasting, as no recent data is available. 

\subsubsection{Commercial buildings in Belgium}

Finally, we also consider PV production data from two commercial buildings in Genk, Belgium. Training data is available for the full year of 2019 for one location, and for nine months in the other location (the entire year of 2020 is available for testing, which will again be univariate). In this case, the exact location of the systems is well-known, but the tilt and azimuth of these systems are not well-specified, because individual panels at the same site are placed in a variety of configurations. The systems have a peak power of 275 kWp and 125 kWp.

\subsection{Source domain}
The source domain data is derived from PVGIS, which uses two methods for calculating PV power data; satellite imagery and climate reanalysis data. The satellite images originate from the METEOSAT geostationary satellites, covering Africa, Europe and parts of Asia. From these satellite images, irradiation at ground level is derived using algorithms developed by CM SAF, creating the SARAH-2 dataset. For locations that do not have satellite data, climate reanalysis data is used, produced by the European Centre for Medium-range Weather Forecast (ECMWF), creating the ERA-5 data set. 

For this research we use data generated by PVGIS from 2005 to 2018, mimicking the parameters (latitude, longitude, tilt, azimuth, peak power) of each of the previously discussed solar PV systems (in case these parameters are available) in order to maximize the similarity between source and target domain. So, for example, an actual PV system in EnergyVille, Genk, Belgium with latitude 50.99 and longitude 5.54 a tilt of 33°, point directly south with a peak power of 2.5 kWp would get these exact parameters in the PVGIS tool to generate data from 2005 until 2018. 

This data is subsequently used to train the source models we use for transfer learning. These models are then fine-tuned using the sparse observational target data. Even though PVGIS provides data for 2020, we limit ourselves to 2018 while creating source models as otherwise this causes data leakage. Data leakage refers to including knowledge about the future (or data which would realistically not be available) in a model. In this case, data leakage would occur if we were to use the years 2019 and 2020 in the creation of our model and, simultaneously, evaluate the model on 2019 and 2020. Similarly, for the Australian PV systems, for which we have data from 2011 to 2012, we limit the source data until 2010. 

\subsection{Weather data}
\label{weather_data_section}

When constructing a model for predicting PV energy generation, weather data can be instrumental for accurate forecasting performance. Variables such as humidity, temperature, cloud cover, and irradiance are highly correlated with the PV power output. Ideally, one would use future weather covariates as these are best aligned with the PV energy generation. In other words, when forecasting the next 24 hours of PV, we would want to include a forecast of any of the aforementioned weather variables. 

The main issue however, is access to historical weather forecasts. While historical weather data is readily available, historical forecasts are more difficult to find. They also introduce an additional dimension to the forecasting problem: the reference point at which the forecast is made. These two issues make it complex to include forecast weather data in deep learning models. 

To mitigate this issue, we treat the source and target domain differently. For training the source model, we utilize readily available observed historical weather data. For the target models, we only utilize actual forecast weather data. While this introduces a misalignment between source and target models, it does not cause any data leakage issues and also avoids placing large and potentially expensive forecast availability requirements on SolNet users, or anyone looking to use transfer-learning for their solar power forecasts.

In this study, we retrieve historical weather data from Open-Meteo via its API. The variables used in this study are relative humidity, direct irradiance, and diffuse irradiance. All that is required is to specify a date-range (which is 2005 until 2019 for our source model), and a latitude and longitude. We did find some open-source historical weather forecasts, but only from 2020 onwards, and only for Europe. The data we use comes from DWD ICON.\footnote{More details can be found on the website of the DWD: https://www.dwd.de/EN/Home/home\_node.html} They maintain both a global and a European Numerical Weather Prediction (NWP) model and make the data publicly available. However, the forecasts are only kept available for a 24 hour period, but Open Climate Fix maintains a repository on Hugging Face for the EU data going back until 2020 \citep{open_climate_fix_2023}. This data is used to fine-tune the transfer model on our target domain and do inference in the target domain. In doing so we can show that using historical weather to train our source models is sufficient to gain a good increase in performance, allowing people to use deep learned models including weather data, which is still easily accessible (through PVGIS and Open-Meteo).

\subsection{Data pre-processing}

The source data is split into a train and test set, while the target data is split into a train set, a validation set, and a separate test set. We do this three-way split to mimic a realistic scenario where a user of the system (e.g. a forecaster) will only have limited data (the train and validation set). So we cannot validate our models during training on this separate test set. We can however use this test set of one year to do a final evaluation of the fully trained model. We split the remaining data in a train and validation set, at a proportion of 80/20.

To train neural networks, our data should also be scaled to a similar range to accelerate convergence of the parameter estimation process. Given that all our variables are non-negative the data is scaled to be between 0 and 1 using a simple min-max scaler (see equation (\ref{eq:scaling})). The scaling is fitted and applied on the train set, and also applied on the validation set, using the fit of the train set, to avoid any data leakage.

\begin{equation}
\label{eq:scaling}
f_{\mathrm{scaled}}(x) = \frac{x - x_{\min}}{x_{\max} - x_{\min}}
\end{equation}

There is an important caveat to the scaling however. While it is fine to use this type of scaling on the source data, it is not as simple in the target domain if we want to apply transfer learning. The model trained on the source domain data has been scaled on 12 years of data. We can therefore reasonably assume that the out-of-sample data will not, on average, be much different from the in-sample data. However, when we naively apply a min-max scaler to the (at times very limited) target domain data, we get data that is, at times, incompatible with the models created in the source domain. For instance, consider the case where we have PV data of a winter month in the Netherlands. When applying a min-max scaler to this unrepresentative dataset, we would scale the values between 0 and 1. But values from summer months would, in that case, go up to 2 or 3, while these values are scaled to 1 in the source model. We tackle this issue by normalizing the data using domain knowledge, i.e. the peak-power (the nominal power on the DC side in kWp) times a factor of one minus 14\% (the typical percentage used by PVGIS to take into account system losses). This peak-power is known in advance for all the systems, avoiding any data leakage while still keeping the data scaled accurately for use in transfer models. Thus, the scaling is still applied as shown in equation (\ref{eq:scaling}), with the addition that $x_{\text{max}}$ for the target data is now:
\[
x_{\text{max}} = \max(x_{\text{max}}, (\text{peak-power} \cdot 0.86))
\]

\section{Results}

In this section, we present results of applying transfer learning to residential PV systems in the Netherlands and Australia, as well as the commercial PV systems in Belgium using univariate data. Next, using the Dutch data, we take a closer look at the impact of including weather covariates. Then we look at the importance of the time of the year when we start gathering target data on model performance. To do this, we change the training period to assess the seasonal impact. Finally, we also assess the impact of data misspecification in the source domain, by increasing the (real) distance between the source and target domain. 

\subsection{Univariate forecasting of residential and commercial PV systems}

The results for univariate forecasting of 40 residential PV systems in the Netherlands are summarised in figure \ref{main_results_NL} (i.e. without incorporating weather observations or forecasts in either the source or target domain). On the x-axis we plot the performance based on the months of data available in the target domain. "0m" is the case with no target data (cf. \ref{fig_performance}. Likewise, "1m" means that one month of target domain data is used for training/finetuning. On the y-axis we plot the scaled RMSE for each of these cases (i.e. the RMSE on the data which has been scaled using the min-max scaler). The thick lines are the overall median\footnote{The median values are shown instead of mean values because not all datasets have the same observation period, which greatly impacts the mean estimate.} values and the shaded areas represent the range of the individual cases. In other words, the shaded area represents the results of all 40 PV systems, from lowest to highest RMSE. The error of the baseline model (a seasonal naive model taking the values of the 24 hours of the previous day to be the new prediction), is a straight line, which does not improve over time, as it is independent of the amount of data available. Finally, identical figures for the MAE and MBE can be found in figures \ref{supp_results_residential} in the appendix.

\begin{figure}[htbp]
    \centering
    \begin{subfigure}[b]{0.48\textwidth}
        \centering
        \includegraphics[width=\textwidth]{Figures/output_NL.pdf}
        \caption{NL scaled RMSE}
        \label{main_results_NL}
    \end{subfigure}
    \hfill
    \begin{subfigure}[b]{0.48\textwidth}
        \centering
        \includegraphics[width=\textwidth]{Figures/output_AUS.pdf}
        \caption{AUS scaled RMSE}
        \label{main_results_AUS}
    \end{subfigure}
    \centering
    \begin{subfigure}[b]{0.48\textwidth}
        \centering
        \includegraphics[width=\textwidth]{Figures/output_BE.pdf}
        \caption{BE scaled RMSE}
        \label{main_results_BE}
    \end{subfigure}
    \caption{Forecast error as a function of target data availability and use of transfer learning for univariate models. The thick line represents the median performance at each time step. The shaded areas represent the performance of all individual systems, with the outer edges being the worst/best performing systems. The Results indicate transfer learning models consistently outperform both naive baselines and target models (i.e. trained without transfer).
    }
    \label{main_results_residential}
\end{figure}

From this plot, we can make four observations. First, the transfer models strictly outperform the baseline models. Second, the target model only starts to outperform the simple baseline once a sufficient amount of data (approx. 4 to 6 months of data depending on the test site) is available to train the model. This is very likely driven by the fact that this time period is sufficient for the model to have seen months with different weather conditions allowing it to generalize. 
Third, we can see that the asymptotic performance improvement is very small, but persistent, after c. 7 months. Fourth, the transfer model almost always outperforms the target model, until more than a half year of data is available. 

Figure \ref{main_results_AUS} plots the corresponding results for Australia. Given the larger size of the dataset (299 samples), the shaded area shows the 0.05 to 0.95 percentiles for each month to filter out some of the noise in the outliers. In a similar fashion to the Dutch results, most of the value lies in the initial months. As before, it takes over 7 months for the target model to consistently outperform the naive model here. The learning improvement performance for the transfer model is again small. The median RMSE is 0.135 for the model with no fine-tuning (0m), improving by about 7.4\% to 0.125 after 12 months. The target models outperform the transfer models in 8.6\% of cases, mostly in the final four months. Interestingly, the asymptotic out-performance is (almost) non-existent for Australia. This could be due to the fact that the source domain is a lot more limited, including only 2005 up until 2010, compared to 2005 up until 2018 for NL.

Figure \ref{main_results_BE} plots the results for the two commercial buildings in Belgium, showing that results are in line with the residential results. As before, the transfer model outperforms the target model, with most of the performance difference in the first six months. The transfer forecasting accuracy is slightly better compared to the Dutch case. This is clear from the baseline RMSEs, which are 0.108 and 0.116 compared to the median baseline of 0.141 in the Netherlands. This is most likely due to the larger scale of the PV installation, as well as the diverse angles of the panels. This causes the output to be smoother compared to smaller installations with one specific angle. 

Finally, it's worth discussing the MAE and MBE of NL, AUS and BE as these metrics tell a more nuanced story than the RMSE. The figures for these metrics can be found in the appendix (\ref{supp_results_residential}). We can see that the MAE of the target and transfer models is worse, relative to the benchmark, than the RMSE. For NL and BE the MAE of the transfer model is on par, and for AUS it is worse than the benchmark. It is worth noting that the target and transfer models are trained based on the MSE criterion, thus favoring a better RMSE. A good MAE is therefore a byproduct of model training, not the main goal. Still, ideally a transfer learned model outperforms the benchmark both on RMSE and MAE. This does become the case if we include weather data, which we discuss in the next section. In terms of the MBE, we can see that target models have a strong positive bias until sufficient data becomes available (3 to 5 months of data), the transfer models however do not show a strong bias. 

\subsection{The effect of weather covariates}

Figure \ref{weather_results_NL} shows the results for the Dutch buildings, but now with the addition of three weather variables: relative humidity at 2 meters, direct solar radiation, and diffuse solar radiation. From the plot, we can come to several conclusions. First, the performance increases significantly when including weather variables, the median transfer error decreases from c. 0.117 to 0.085, a performance increase of 27.4\%. Second, the performance increase is even more pronounced with target models. When no covariates were included, 7 months of data were needed by the target LSTM to outperform even the naive model on average, but with weather data it only takes 2 months. On average, the transfer model still performs better for the first half year, but the difference is smaller compared to the univariate case, and the asymptotic out-performance is also more limited. A final observation is that the performance is far more consistent across different sites now, i.e. the confidence band of the target and transfer models is a lot narrower now. 

\begin{figure}[!t]
\centering
\includegraphics[width=4.5in]{./Figures/weather_results.pdf}
\caption{Forecast error as a function of target data availability and use of transfer learning for multivariate models using weather data. The target model is trained on weather forecast data, the transfer model is trained on actuals weather data and fine-tuned on weather forecast data. Results are consistent with the univariate models, but with a strong increase in overall performance.}
\label{weather_results_NL}
\end{figure}

When we look at the MAE and MBE of the models that include weather data (see figures \ref{supp_results_weather}), the story also changes significantly compared to the models without weather data. When weather data is included as a predictor, we can see that the MAE is in line with the RMSE. The transfer models consistently outperform the benchmark on the MAE, and the target model does so after sufficient data is available to train on (after c. 3 to 5 months). The univariate transfer models thus seem to be adequate in finding the overall shape of the solar curve, limiting the large errors, while the weather data adds in the small errors that happen hour-to-hour. Finally, we can see that the bias in the target models is also gone much faster when weather data is included, while it has no real effect on the bias in transfer models (which was already low).  

\subsection{The effect of seasonality}
\label{seasonality}

Next, we focus on the aspect of yearly seasonality. Given that the target models catch up to the transfer learning performance after over half a year, it is reasonable to expect that the learner requires some of the summer (or transition) months to be trained on to achieve decent performance. The training so far was done chronologically with July as the final month for NL and AUS, and December as the final month for BE, i.e. for the case of '3m' or three months of data, the data comes from the months of October, November and December for BE. The training and validation split was also done chronologically, i.e. the validation data for this '3m' case consists of the last 18 days of December. 

To explore this further, we shifted the terminal month in time for different, univariate models. Table \ref{tab_seasonal} shows these results for NL, by focusing specifically on what happens with the RMSE of the target models relative to the daily persistence model (giving an RMSE skill-score). The columns represent the final month of the target data used for training. The earlier in the year 2019 we have the terminal month, the less data we have in total, hence the increasing amount of n.a. values. We can see that there is a clear trend: the earlier in the year the final month is, the better, up until a point. This most likely has to do with the presence of the best performing day inside the training data. Out of the 40 datasets, the month with the highest PV-output hour was in April 21 times and in May eleven times. As soon as this data is available to the model, the performance drastically increases. It is also the case that the main forecasting errors come from underestimating the peaks, which further explains the need for this data to be available during the training phase.

\begin{table}[!t]
\centering
\caption{RMSE of target model for NL relative to the baseline dependent on the terminal training month (the columns) and the amount of target data to train and validate on (the rows)}
\label{tab_seasonal}
\fontsize{7}{8}\selectfont
\setlength{\tabcolsep}{6pt} 
\renewcommand{\arraystretch}{1.2} 

\begin{tabular}{|>{\centering\arraybackslash}p{0.7cm}|*{12}{>{\centering\arraybackslash}p{0.85cm}|}}\hline
\rowcolor{lightgray}
\multicolumn{1}{|c|}{} & \multicolumn{12}{c|}{\textbf{Final Month}} \\
\hline
\rowcolor{lightgray}
& \fontsize{6}{7}\selectfont\textbf{Jan/20} & \fontsize{6}{7}\selectfont\textbf{Dec/19} & \fontsize{6}{7}\selectfont\textbf{Nov/19} & \fontsize{6}{7}\selectfont\textbf{Oct/19} & \fontsize{6}{7}\selectfont\textbf{Sep/19} & \fontsize{6}{7}\selectfont\textbf{Aug/19} & \fontsize{6}{7}\selectfont\textbf{Jul/19} & \fontsize{6}{7}\selectfont\textbf{Jun/19} & \fontsize{6}{7}\selectfont\textbf{May/19} & \fontsize{6}{7}\selectfont\textbf{Apr/19} & \fontsize{6}{7}\selectfont\textbf{Mar/19} & \fontsize{6}{7}\selectfont\textbf{Feb/19} \\
\hline
0m  & 48\% & 47\% & 47\% & 47\% & 47\% & 47\% & 47\% & 47\% & 47\% & 47\% & 47\% & 47\% \\
1m  & 62\% & 58\% & 53\% & 31\% & 7\% & 5\% & 13\% & 31\% & 6\% & 7\% & 14\% & 20\% \\
2m  & 58\% & 52\% & 35\% & \cellcolor{mediumgreen} -3\% & 0\% & 5\% & 6\% & 20\% & 1\% & \cellcolor{mediumgreen2} -8\% & 0\% & n.a. \\
3m  & 58\% & 23\% & 0\% & \cellcolor{mediumgreen2} -9\% & 2\% & 10\% & 7\% & 8\% & \cellcolor{mediumgreen2} -8\% & \cellcolor{mediumgreen2} -9\% & n.a. & n.a. \\
4m  & 22\% & \cellcolor{mediumgreen2} -8\% & \cellcolor{mediumgreen2} -10\% & \cellcolor{mediumgreen2} -8\% & 5\% & 13\% & 5\% & 4\% & \cellcolor{darkgreen} -11\% & n.a. & n.a. & n.a. \\
5m  & \cellcolor{mediumgreen} -7\% & \cellcolor{darkgreen} -11\% & \cellcolor{darkgreen} -11\% & \cellcolor{lightgreen} -1\% & 9\% & 7\% & \cellcolor{mediumgreen} -4\% & \cellcolor{darkgreen} -11\% & n.a. & n.a.& n.a.& n.a. \\
6m  & \cellcolor{darkgreen} -13\% & \cellcolor{darkgreen} -12\% & \cellcolor{darkgreen} -11\% & 6\% & 6\% & \cellcolor{mediumgreen} -4\% & \cellcolor{mediumgreen2} -9\% & n.a. & n.a. & n.a. & n.a. & n.a. \\
7m  & \cellcolor{darkgreen} -14\% & \cellcolor{darkgreen} -12\% & \cellcolor{mediumgreen2} -10\% & 6\% & \cellcolor{lightgreen} -2\% & \cellcolor{mediumgreen2} -9\% & n.a. & n.a. & n.a. & n.a. & n.a. & n.a. \\
8m  & \cellcolor{darkgreen} -13\% & \cellcolor{darkgreen} -11\% & \cellcolor{darkgreen} -12\% & 1\% & \cellcolor{mediumgreen2} -7\% & n.a. & n.a. & n.a. & n.a.& n.a. & n.a. & n.a. \\
9m  & \cellcolor{darkgreen} -13\% & \cellcolor{mediumgreen2} -9\% & \cellcolor{mediumgreen} -4\% & \cellcolor{mediumgreen2} -7\% & n.a. & n.a. & n.a. & n.a. & n.a.& n.a. & n.a. & n.a. \\
10m & \cellcolor{darkgreen} -14\% & \cellcolor{mediumgreen2} -9\% & \cellcolor{mediumgreen2} -8\% & n.a. & n.a. & n.a. & n.a. & n.a. & n.a. & n.a. & n.a. & n.a.\\
11m & \cellcolor{darkgreen} -13\% & \cellcolor{darkgreen} -11\% & n.a. & n.a. & n.a. & n.a. & n.a. & n.a. & n.a. & n.a. & n.a. & n.a. \\
12m & \cellcolor{darkgreen} -11\% & n.a. & n.a. & n.a. & n.a. & n.a. & n.a. & n.a. & n.a. & n.a. & n.a. & n.a. \\
\hline
\end{tabular}
\end{table}

\subsection{The effect of source data misspecification}
\label{chap:quan_and_qual}

As a final step to test the generalisation of the model, we focused on the role of the source domain and the quality of the pre-trained model. The experimental setup proposed in this paper is unique in this aspect, as pre-training on prior observed data from other similar sites does not necessarily provide this flexibility. More specifically, we assess the impact of increasing the distance between the source and target domain by using source data increasingly to the south-east of the original sites (at an angle of 125° to avoid ending up in the Adriatic Sea after 800km). Figure \ref{mis_specification} plots these results for a subset of ten locations of our NL dataset. Again, the lines show the median values, while the shaded area shows the confidence interval. 

Two key findings become clear from these figures. On the one hand, we can see that misspecification of the target location comes with increasingly erroneous forecasts. On the other hand, the use of external covariates plays a key role in the impact of these misspecifications. When weather covariates are included in the model (Figure \ref{weather_mis}), the results worsen drastically with increased distance from the target location. Moving 100km to the south-east increases the error by 15\% (from 0.080 to 0.093) over the correct location, 35\% at 200km (from 0.080 to 0.108) and at 400km the error is already at a similar level to a naive model. When no weather variables are included (Figure \ref{regular_mis}), this effect is far less pronounced. The performance definitely lags behind on a model with weather covariates, but the model is much more robust towards misspecifications. The effect on the MAE (figure \ref{supp_results_incorrect}) is similar to that of the RMSE. Weather data adds in predictive performance, but has a detrimental effect when the target domain is misspecified.

The median target RMSE (the blue line) also makes clear how important correct specification of the target location is when weather data is included. When the location is misspecified by c. 200km we can see that the transfer model has already lost all its relative value after two months. When the misspecification is as small as 100km, the value is lost after 5 months. However, when no weather data is included, a misspecification of the source domain has almost no effect on target outperformance, even after more than 6 months.

\begin{figure}[htbp]
    \centering
    \begin{subfigure}[b]{0.48\textwidth}
        \centering
        \includegraphics[width=\textwidth]{Figures/incorrect_no_weather.pdf}
        \caption{Without weather covariates}
        \label{regular_mis}
    \end{subfigure}
    \hfill
    \begin{subfigure}[b]{0.48\textwidth}
        \centering
        \includegraphics[width=\textwidth]{Figures/incorrect_weather.pdf}
        \caption{With weather covariates}
        \label{weather_mis}
    \end{subfigure}
    \caption{Transfer learning results under increasing distance from the target for a subset of ten NL locations. The performance is compared to the median baseline of these ten locations, as well as the median 'correct' target model. I.e. a target model that is made with no misspecification.}
    \label{mis_specification}
\end{figure}

\section{Discussion}

The findings in the result section are similar across all locations, with most of the performance gain offered by transfer learning, using the synthetic data, coming from the difference in initial performance. This provides for a useful tool for any PV installation with limited to no data available. Using only open-source historical PV and weather data, and ideally some meta-data on the actual PV installations, a model can be made to give accurate zero-shot PV predictions anywhere in the world. 

There are several important considerations to be made to assess the value of transfer learning. Weather data has a strong positive effect on the predictive performance of models, and the translation from actuals in the source model as a proxy for forecasts, to actual forecasts holds up. If no data is available from a sunnier time period, the transfer model will have considerably more added value that will last longer, although here as well, weather data in the transfer model mitigates this effect. This indicates that most of the difficulty is in correctly learning to forecast the seasonality inherent in PV energy production and, more importantly, the peak production. Finally, it is important to try to mimic the situation of the target data (location, azimuth, slope) to limit the distance between source and target domain. However, the transfer models do appear to be robust to small misspecifications, but larger misspecifications can have a detrimental effect, especially when detailed external covariates, tied with the location, are used for predictive purposes.

\subsection{Future directions and limitations}

In this paper, we have considered the LSTM model as the neural architecture to train the solar power forecasters. While other, more recent deep learning architectures exist (e.g. transformer models, NBeats, Nhits) that can potentially achieve even better results, the goal of this paper was to provide a principled study into transfer learning as a tool to improve performance compared to simple statistical benchmarks on the one hand while simultaneously improving sample complexity of deep neural networks. We expect the results to generalize to these other neural architectures, but this is the subject of a follow-up study.

One limitation of the paper is the lack of access to datasets from climate zones outside of the temperate (maritime) climate zones of the Netherlands, Belgium, and even Australia, where all datasets originate from New South Wales. To further substantiate the added value of SolNet, the results shown in this study should be replicated in other parts of the world. We expect that, by making the code used for this paper freely available, more datasets will be used to validate the global performance of transfer models. This will also allow other researchers to build on top of the models proposed here to achieve the same multiplier effect as seen in other domains such as computer vision etc. 

\section{Conclusion}
\label{conclusion}

This paper provides a detailed assessment of the impact of utilising a large corpus of open-source (synthetic) data in the creation of data-driven forecast models for solar power output. The improvements demonstrated hold regardless of the location of the PV panels, and are rather robust to misspecifications in system parameters. As evidence, the paper looked at over 300 sites in three different countries to substantiate the findings while running extensive tests lasting one year on all sites.

This methodology extends existing recent work on transfer learning by minimising the need for data from 'similar sites', and shows that there is a large boost to the initial predictive performance. As expected, this differential subsides over time, and is almost zero after a one year period, opposed to what we hypothesised (see figure \ref{fig_performance}). But the impact of using transfer models is accumulative over time and dependent on several other factors like seasonality, giving users a powerful tool and a set of guidelines to improve their forecasting performance at the early onset of (or even before) installing a new PV setup. 

The code, along with a functioning example for using SolNet for solar forecasting anywhere on land around the globe, has been open-sourced\footnote{https://github.com/Joris-Dprtr/SolNet}. We expect this work to be the first of many that have the potential to lead to general-purpose deep learning models for energy forecasting.


\newpage
\appendix
\section{Supplementary figures}

\begin{figure}[htbp]
    \centering
    \begin{subfigure}[b]{0.48\textwidth}
        \centering
        \includegraphics[width=\textwidth]{Figures/output_NL_MAE.pdf}
        \caption{NL scaled MAE}
        \label{main_results_NL_MAE}
    \end{subfigure}
    \begin{subfigure}[b]{0.48\textwidth}
        \centering
        \includegraphics[width=\textwidth]{Figures/output_NL_MBE.pdf}
        \caption{NL scaled MBE}
        \label{main_results_NL_MBE}
    \end{subfigure}
    \centering
    \begin{subfigure}[b]{0.48\textwidth}
        \centering
        \includegraphics[width=\textwidth]{Figures/output_AUS_MAE.pdf}
        \caption{AUS scaled MAE}
        \label{main_results_AUS_MAE}
    \end{subfigure}
        \centering
    \begin{subfigure}[b]{0.48\textwidth}
        \centering
        \includegraphics[width=\textwidth]{Figures/output_AUS_MBE.pdf}
        \caption{AUS scaled MBE}
        \label{main_results_AUS_MBE}
    \end{subfigure}
        \centering
    \begin{subfigure}[b]{0.48\textwidth}
        \centering
        \includegraphics[width=\textwidth]{Figures/output_BE_MAE.pdf}
        \caption{BE scaled MAE}
        \label{main_results_BE_MAE}
    \end{subfigure}
        \centering
    \begin{subfigure}[b]{0.48\textwidth}
        \centering
        \includegraphics[width=\textwidth]{Figures/output_BE_MBE.pdf}
        \caption{BE scaled MBE}
        \label{main_results_BE_MBE}
    \end{subfigure}
    \caption{The mean absolute error (MAE) and mean biased error (MBE) from the base experiments. Compared to the RMSE, the MAE of the target and transfer models is notably worse relative to the baseline. In terms of bias, a target made model suffers from positive bias until several months of data are included in the training process.
    }
    \label{supp_results_residential}
\end{figure}

\begin{figure}[htbp]
    \centering
    \begin{subfigure}[b]{0.48\textwidth}
        \centering
        \includegraphics[width=\textwidth]{Figures/output_NL_Weather_MAE.pdf}
        \caption{NL scaled MAE including weather variables}
        \label{weather_results_NL_MAE}
    \end{subfigure}
    \hfill
    \begin{subfigure}[b]{0.48\textwidth}
        \centering
        \includegraphics[width=\textwidth]{Figures/output_NL_Weather_MBE.pdf}
        \caption{NL scaled MBE including weather variables}
        \label{weather_results_NL_MBE}
    \end{subfigure}

    \caption{The mean absolute error (MAE) and mean biased error (MBE) from the experiments including weather data. Once weather data is included in the models, the MAE follows the same trend as the RMSE, with the transfer and the target model (eventually) outperforming the benchmark. The bias is also gone faster in a target model once weather data is included. 
    }
    \label{supp_results_weather}
\end{figure}

\begin{figure}[htbp]
    \centering
    \begin{subfigure}[b]{0.48\textwidth}
        \centering
        \includegraphics[width=\textwidth]{Figures/output_NL_incorrect_MAE.pdf}
        \caption{NL scaled MAE with increasing distance from the target}
        \label{incorrect_results_NL_MAE}
    \end{subfigure}
    \hfill
    \begin{subfigure}[b]{0.48\textwidth}
        \centering
        \includegraphics[width=\textwidth]{Figures/output_NL_incorrect_MBE.pdf}
        \caption{NL scaled MBE with increasing distance from the target}
        \label{incorrect_results_NL_MBE}
    \end{subfigure}
    \centering
    \begin{subfigure}[b]{0.48\textwidth}
        \centering
        \includegraphics[width=\textwidth]{Figures/output_NL_incorrect_weather_MAE.pdf}
        \caption{NL scaled MAE with increasing distance from the target including weather variables}
        \label{incorrect_weather_results_NL_MAE}
    \end{subfigure}
    \hfill
    \begin{subfigure}[b]{0.48\textwidth}
        \centering
        \includegraphics[width=\textwidth]{Figures/output_NL_incorrect_weather_MBE.pdf}
        \caption{NL scaled MBE with increasing distance from the target including weather variables}
        \label{incorrect_weather_results_NL_MBE}
    \end{subfigure}
    \caption{The mean absolute error (MAE) and mean biased error (MBE) from the experiments with increasing misspecification of the target domain. The top figures do not have weather data included and the bottom figures do have weather data included. Note that the figures without weather data no longer align with \ref{supp_results_residential} (which is clear from the negative bias that can now be seen in the target model in \ref{incorrect_results_NL_MBE}) as different months are used in order to compare the results with the results including weather data (for which only 8 months are available, instead of 12). 
    }
    \label{supp_results_incorrect}
\end{figure}

\end{document}